\documentclass[12pt]{article}
\usepackage{latexsym, epsfig, graphics, amsmath,}
\numberwithin{equation}{section}
\newcommand{\be}{\begin{equation}}
\newcommand{\ee}{\end{equation}}
\newcommand{\bq}{\begin{eqnarray}}
\newcommand{\eq}{\end{eqnarray}}

\newcommand{\rhz}{\rho_{0}}
\newcommand{\lmz}{\lambda_{0}}
\newcommand{\eyz}{A_{0}(q)}
\newcommand{\eyzs}{A_{0}^{\star}(q)}
\newcommand{\mzs}{m_{0}^{2}}
\newcommand{\vts}{v(\tau,\sigma)}

\newcommand{\mz}{m_{0}}
\newcommand{\bsi}{\bar{\sigma}}
\begin{document}
\begin{titlepage}
\today          \hfill 
\begin{center}

%\hfill    LBNL-xxxxx \\
%          \hfill    UCB-PTH-xx/xx \\

\vskip .5in

{\large \bf QCD\,3 On The World Sheet}
\footnote{This work was supported 
 by the Director, Office of Science,
 Office of High Energy  
 of the U.S. Department of Energy under Contract No.
DE-AC02-05CH11231}.
\vskip .50in
K.Bardakci \footnote{Email: kbardakci@lbl.gov}

{\em Theoretical Physics Group\\
    Lawrence Berkeley National Laboratory\\
      University of California\\
    Berkeley, California 94720}
\end{center}

\vskip .5in

\begin{abstract}
In this article, we apply the world sheet approach developed in
earlier work to QCD in 1+2 dimensions. The starting point is a
field theory on the world sheet that reproduces the planar graphs
of QCD parametrized by the light cone variables. We study 
the ground state
of this model using a variational approximation, and show that it 
consists of a set graphs infinitely dense on the world sheet. The
energy of this new ground state is lower than  that of the empty
world sheet, signaling a phase transition. Also, a finite mass is
generated in the originally massless theory. Finally, we consider
a particular set of time dependent fluctuations about the static
ground state configuration, which result in the formation of a string
on the world sheet.

\end{abstract}
\end{titlepage}%THIS PAGE (PAGE ii) CONTAINS THE LBL DISCLAIMER

\newpage
\renewcommand{\thepage}{\arabic{page}}
\setcounter{page}{1}
%THIS IS PAGE 1 (INSERT TEXT OF REPORT HERE)
\section{Introduction}
\vskip 9pt

The present article is the continuation of a series of previous
articles [1, 2], based on the program of the world sheet description of
planar graphs of field theories. The field theory studied in the
previous work was the $\phi^{3}$ theory, chosen for its simplicity.
In the present paper, we apply the methods developed in the earlier
work to the more interesting and more physical $QCD\,3$ in
$1+2$ dimensions. This model has been studied in the literature
extensively using various different approaches [3]. The world 
sheet formulation of $QCD$ we are going to use here was developed in
[4,5,6].

The goal of the program is to sum the planar graphs of a field
theory on the world sheet parametrized by the light cone
variables [7]. It was shown in [8] that this sum is 
reproduced by a two dimensional field theory that lives on
the world sheet.
The world sheet field theories for $\phi^{3}$ as well as for
$QCD\,3$ are non-local and
 somewhat complicated. The challenge is then to find a manageable
approximation scheme which captures the essence of the model. The
scheme we use in this paper is the well-known variational approach
for determining the ground state of a Hamiltonian. With the help
of a suitably chosen trial state, we establish
the following results for $QCD\,3$:\\

 The ground state of the theory consists of a set of light cone graphs
infinitely dense on the world sheet. This is clearly non-perturbative,
and in fact, it signals a phase transition triggered by the
condensation of the graphs. We also show that in this new phase,
the originally massless theory develops a finite mass.

The ground state is a static (solitonic) configuration. Next, we
introduce a particular set of time dependent fluctuations around this
static background. These fluctuations restore translation invariance
on the world sheet violated by the static background, and they also
generate a string on the world sheet. We show this by computing the
action that describes these modes.

The following is a  preview of the  sections of this paper. In section
2, the rules for light cone graphs on the world sheet are reviewed,
and the world sheet which reproduces these graphs is described in
section 3. In section 4, the trial state for the variational
calculation is introduced. This state is simple enough to allow 
explicit calculations, and it also avoids the divergences peculiar
to the light cone picture.

In section 5, we derive the variational equations for the trial state,
and then partially solve them. The resulting energy of the ground
state has a linear divergence in the integral over the transverse
momentum $q$. This is due to translation invariance in this variable;
 the ground state energy is proportional to the volume
in the $q$ space. We argue that the relevant finite quantity is then
the energy per unit volume.

We note that the ground state energy is negative,  lower than
the energy of  the empty world sheet, which is zero. This signals a
phase transition in the model. The interesting physics is then in the
spectrum of excitations localized in $q$ above the ground state. These
excitations are massive, which means that a finite mass is generated
in the originally massless theory.

So far, we have only only solved the variational equations partially;
the ground state energy has still to be minimized with respect to
two remaining parameters $a$ and $\rhz$. In section 6, we determine
the dependence of the energy on these parameters, and in section 7,
we show that the minimum is reached in the limit $a\rightarrow 1$
annd $\rhz \rightarrow \infty$. The crucial parameter $m_{0}$
that charaterizes the local excitations is finite as 
 $\rhz \rightarrow \infty$, but diverges as $g^{2}/(a-1)$ as
$a\rightarrow 1$. We argue that there is no adjustable coupling
constant in this problem; $g$ merely provides the correct dimension
for the energy. We may then let $m_{0}$ instead of $g$ provide
this dimension and thereby eliminate the singularity in $a-1$.

In closing this section, we compute the density of the light cone
graphs on the world sheet, and find that it goes to $\infty$
as $\rhz \rightarrow \infty$. It is than reasonable to expect
that these dense set of graphs will lead to string formation, a
problem which we investigate in the next section.

String formation is the result of a particular quantum
fluctuation around the original static ground state configuration [1,2].
It corresponds to shifting the the momentum $q$ by a fluctuating
field $\vts$. This field can be considered as the collective coordinate
which restores the translational symmetry in $q$ broken by the
ground state configuration. It will also turn out to be the string
coordinate.

In the rest of the section, we compute the contribution of this
field to the action, after a somewhat lenghty but straightforward
algebra, the result turns out to be the standard string action
in the light cone picture. There is, however, a problem: The constant
in front of the kinetic energy term is singular in the limit
$a \rightarrow 1$ and $\rhz \rightarrow \infty$. This singularity can
be eliminated by a suitable scaling of both the coordinate $\sigma$
and the field $v$. After this scaling, the original $\sigma$
interval of finite length becomes an interval of infinite length.
We interpret this as having an action for the bulk of the string,
leaving the boundary conditions to be fixed at the end. One can then
supply the boundary conditions by cutting a finite piece of
the string and glueing the end points together. We should, however,
emphasize that the emerging standard string picture is only an
approximation, since only the fluctuation of the momentum $q$
was taken into account. The parameters $a$ and $m_{0}$ were fixed
and their fluctuations were neglected. Their inclusion would lead
to a more complicated string picture, which we hope to investigate
in the future.

In the final section, we  summarize our results and discuss directions
for future research.

\section{The World Sheet Picture}
\vskip 9pt

The planar graphs of the free part of $QCD\,3$ are the same as in the massless
scalar $\phi^{3}$ theory. They
can be represented on a world sheet
parametrized by the $\tau=x^{+}$ and $\sigma=p^{+}$ as a collection of
horizontal solid lines (Fig.1), where the n'th line carries the one
dimensional transverse momentum $q_{n}$.
\begin{figure}[t]
\centerline{\epsfig{file=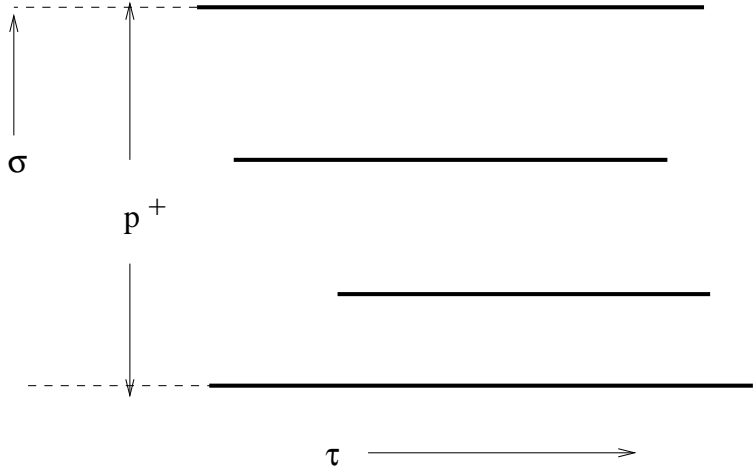, width=10cm}}
\caption{A Typical Graph} 
\end{figure}
Two adjacent solid lines labeled by n and n+1 correspond to the light
cone propagator
\be
\Delta(p_n)=\frac{\theta(\tau)}{2 p^{+}}\,\exp\left(-i
\tau\,\frac{p_{n}^{2} }{2\,p^{+}}\right),
\ee
where $p_n= q_{n+1}- q_{n}$ is the transverse momentum  and
$$
p^{+}_{n}=\sigma_{n+1} -\sigma_{n},
$$
is the light cone momentum flowing through the propagator.

 In the interacting theory, in addition to the 
propagators, there are three and four point vertices. The two three
point vertices are pictured in Fig.2. When lines 1 and 2 merge to form
the line 3, the associated vertex factor is given by
\be
V(1+2\rightarrow 3)=
\left(\frac{\sigma_{2} -\sigma_{1}}{\sigma_{3} -\sigma_{2}}
+\frac{\sigma_{3}-\sigma_{2}}{\sigma_{3}-\sigma_{1}}\right)\,p_{2}
-\left(\frac{\sigma_{2} -\sigma_{1}}{\sigma_{3} -\sigma_{1}} +
\frac{\sigma_{3} -\sigma_{2}}{\sigma_{2} -\sigma_{1}}\right)\,
p_{1}.
\ee

The vertex factor $V(3\rightarrow 1+2)$, for line 1 splitting into
lines 2 and 3, is given by the conjugate expression. We will not write
down the four point vertex since it will not be needed in the present work.

\begin{figure}[t]
\centerline{\epsfig{file=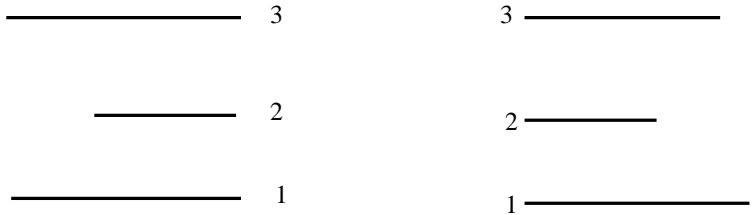, width=10cm}}
\caption{Interaction Vertices}
\end{figure}

\vskip 9pt

\section{The World Sheet Field Theory}

\vskip 9pt

The light cone graphs described above are generated by a world sheet
field theory. We introduce a complex scalar field $\phi(\tau,\sigma,
q)$ and its conjugate $\phi^{\dagger}$, which at time $\tau$,
annihilate (create) a solid line with coordinate $\sigma$, carrying
momentum $q$. They satisfy the usual commutation relations:
\be
[\phi(\tau,\sigma, q), \phi^{\dagger}(\tau,\sigma',q')]=\delta(\sigma-
\sigma')\,\delta(q-q').
\ee
The vacuum, annihilated by the $\phi$'s, represents the empty world
sheet.
For later use, it is also convenient to define the composite operator
$\rho$ which represents the density of the solid lines:
\be
\rho^{2}(\tau,\sigma)=\int
dq\,\phi^{\dagger}(\tau,\sigma,q)\,\phi(\tau,\sigma,q).
\ee

The free Hamiltonian consists of a bunch of solid lines, representing
free propagators. An important restriction is that propagators are
assigned to adjacent solid lines, and not to the non-adjacent ones. To
enforce this constraint, we need to define the projection operator 
$\mathcal{E}(\sigma_{i},\sigma_{j})$. An explicit construction of this
operator in terms of auxilliary
fermionic fields was given in [1]. However, this construction is not
needed in the present case. All we need to know is the action of this
projection operator on the states $|s\rangle$ that will be defined in
the next section. This action is defined by the equations
$$
\mathcal{E}(\sigma_{i},\sigma_{j})|s\rangle =0
$$
if $\sigma_{j}\leq \sigma_{i}$.
$$
\mathcal{E}(\sigma_{i},\sigma_{j})|s\rangle =0
$$
if $\sigma_{j}>\sigma_{i}$ \underline{and} there are solid lines
between $\sigma=\sigma_{i}$ and $\sigma=\sigma_{j}$.
$$
\mathcal{E}(\sigma_{i},\sigma_{j})|s\rangle=|s\rangle
$$
if  $\sigma_{j}>\sigma_{i}$ \underline{and} there are no solid lines
between  $\sigma=\sigma_{i}$ and $\sigma=\sigma_{j}$. These equations
are all that is needed to compute the matrix elements
$\langle s|\mathcal{E}(\sigma_{i},\sigma_{j})|s\rangle $ and derive
equations (6.9), (6.15) and (8.8).
Also, using the  properties of the projection operator described above,
 the free Hamiltonian
can be written as
\bq
H_{0}&=&\frac{1}{2} \int d\sigma \int d\sigma' \int dq \int dq'\,\frac{
\mathcal{E}(\sigma,\sigma')}{\sigma' -\sigma}\, (q-q')^{2}\nonumber\\
&\times& \phi^{\dagger}(\sigma,q)\phi(\sigma,q)\,\phi^{\dagger}(\sigma',q')
\phi(\sigma',q')\nonumber\\
&+&\int d\sigma\,\lambda(\sigma) \left(\int dq\,\phi^{\dagger}(\sigma,q)
\phi(\sigma,q) -\rho^{2}
(\sigma)\right),
\eq
where $\lambda$ is a Lagrange multiplier.

The interaction Hamiltonian, which reproduces vertex factors of (2.2), is
given by
\bq
H_{I}&=& ig\,
\int d\sigma_{1} \int d\sigma_{2} \int d\sigma_{3}\,
\theta(\sigma_{2}-\sigma_{1})\,\theta(\sigma_{3}-\sigma_{2})\,
\frac{\mathcal{E}(\sigma_{1},\sigma_{3})}{\sqrt{(\sigma_{2}-\sigma_{1})\,
(\sigma_{3}-\sigma_{2})\,(\sigma_{3}-\sigma_{1})}}\nonumber\\
&\times&\left(1+\frac{\sigma_{3}-\sigma_{2}}{\sigma_{2}-\sigma_{1}}+
\frac{\sigma_{2}-\sigma_{1}}{\sigma_{3}-\sigma_{2}}\right)\,
\rho^{2}(\sigma_{1})
\,\rho^{2}(\sigma_{3})\,\int dq_{2}\,q_{2}\,\phi(\sigma_{2},q_{2})+H.C.
\eq
The $\theta$ functions order the $\sigma$ integrations so that
$\sigma_{1}<\sigma_{2}<\sigma_{3}$. The total Hamiltonian
\be
H=H_{0}+H_{I},
\ee
as well as the commutation relations (3.1), follow from the action
\be
S=\int d\tau\,\left(i\,\int d\sigma \int dq\,\phi^{\dagger}\partial_{\tau}
\phi\,-H(\tau)\right).
\ee

An important feature of this action is its symmetries..
It is invariant 
under the light cone subgroup of Lorentz transformations, and also
under translations of the transverse momentum,
\be
\phi(\tau,\sigma,q)\rightarrow \phi(\tau,\sigma, q +r),
\ee  
by a constant $r$, as well as translations in $\sigma$ and $\tau$
coordinates.
 Among the lightcone symmetries, the boost along
the special direction 1 is of special importance. Under this
transformation, parametrized by $u$, the fields transform as
\be
\phi(\tau,\sigma,q)\rightarrow\sqrt{u}\,\phi(u\tau, u\sigma, q),\,\,
\lambda(\tau,\sigma)\rightarrow u\,\lambda(u\tau,u\sigma),\,\,
p^{+}\rightarrow\frac{1}{u}\,p^{+}.
\ee
To simplify the algebra, we take advantage of this invariance and set,
\be
p^{+}=1,
\ee
by taking $u=p^{+}$. The correct $p^{+}$ dependence can always be
restored at the end of a calculation.

Another important symmetry is
\be
\phi(\tau,\sigma,q)\rightarrow -\phi(\tau, \sigma,-q),\,\,
\phi^{\dagger}(\tau,\sigma,q)\rightarrow -\phi^{\dagger}(\tau,
\sigma,-q).
\ee

These symmetries allow us to simplify the search for the ground state.
We follow the common practice and assume that the ground state
configuration is invariant under these symmeries.

\vskip 9pt
 
\section{The Setup For The Variational Calculation}

\vskip 9pt
 
In the standard variational approach, the approximate ground state
energy and the wave function is computed by sandwiching the
Hamiltonian between suitably chosen trial states and minimizing the
energy with respect to the variational parameters. In our case, an
arbitrary state is generated by applying a product of $\phi^{\dagger}$'s 
at various values of $\sigma$'s and $q$'s but at a fixed value of $\tau$
to the vacuum. In this section, we will introduce the trial state
we will use and carry out part of the variational calculation. The
simplest state often used in problems of this type is the coherent
state
\be
|c\rangle=N^{-1}\,\exp\left(\int d\sigma\,\int dq\,f(\sigma,q)\,
\phi^{\dagger}(\sigma,q)
\right)\,|0\rangle,
\ee
where $f$ is the variational function and N is the normalization
constant. This corresponds to a random distribution of solid lines
with no correlations between them. In our case, there is a problem with
using this state: When two adjacent lines with coordinates
$\sigma_{1}$ and $\sigma_{2}$ approach each other,
$$
\sigma_{2} -\sigma_{1}\rightarrow 0,
$$
the Hamiltonian (3.3, 3.4), becomes singular, and the variational energy
blows up. This can be traced back to the $1/p^{+}$ factors in (2.1),
typical of the light cone picture. The simplest way to avoid this
blowup is to introduce  correlation functions between adjacent solid
lines which vanish sufficently fast as the lines approach each
other. This is a general feature of quantum mechanics: At the locations
where the potential becomes singular, the wave function has to vanish.
Accordingly, we define the auxiliary states by the recursion relation
\be
|n+1,\sigma\rangle=K(\sigma)\,\int_{0}^{\sigma}
d\sigma'\,g(\sigma-\sigma')
|n,\sigma'\rangle,
\ee
and the initial condition
\be
|n=0,\sigma\rangle=|0\rangle,
\ee
where $n$ is a positive integer and $\sigma$ ranges from $0$ to
$p^{+}=1$. 

 For the correlation function $g$, we choose
\be
g(\sigma-\sigma')=(\sigma-\sigma')^{a/2},
\ee
where $a$ is a positive variational parameter. For sufficently large
$a$, there is no energy blowup. This choice is motivated by the simplicity
of this ansatz, which makes it possible to compute various quantities
 explicitly.

We take  for $K$
\be
K(\sigma)=\int dq\,A(\sigma,q)\,\phi^{\dagger}(\sigma,q),
\ee
and  the variational state is then given by
\be
|s\rangle=\sum_{n=2}^{\infty}|n,\sigma=1\rangle.
\ee
We note that in this ansatz, the dependence on $q$ and $\sigma$
factorizes. It is then easy to show that the contribution of the four
point vertex vanishes. This is, of course, a feature of this particular
ansatz and is not true in general.

This variational state is closely related to the coherent state.
In fact, for
$$
a=0,
$$
it is  the coherent state with
$$
f(\sigma,q)=A(\sigma,q).
$$
We also note that sum over $n$ starts at $n=2$ instead of $n=0$, so
that, empty world sheet and a single line are eliminated.

We now have to compute the normalized expectation value of the
Hamiltonian,
\be
\langle H \rangle\equiv N^{-1}\, \langle s|H|s \rangle,
\ee
as a function of $A$, $\rho$ and $\lambda$, the variational parameters
of the problem, and  solve the corresponding variational
equations. We designate the solutions to these equations by a
subscript $0$, and take $\eyz$, $\lmz$ and $\rhz$ to
be independent of
$\sigma$. This is because the ground state wavefunction is expected to
be invariant under the symmetries of the problem, in this case,
translation invariance in $\sigma$. We will discuss invariance under
translations of $q$ later on. From now on, we will use the notation
$$
\langle O \rangle=N^{-1}\,\langle s|O|s \rangle
$$
for the normalized expectation value of any operator $O$.

 In the next section, we will  solve the variational equation 
  for $A$, and leave the
rest to the subsequent sections.

\vskip 9pt

\section{The Variational Equation For A}

\vskip 9pt

By sandwiching $H$ between the states $|s \rangle$, it is easy to show
that the normalized expectation values of the various terms of the
Hamiltonian are of the form
\bq
\langle H_{0} \rangle&=&Z_{0}\,\int dq\,q^{2}\, |A(q)|^{2},\nonumber\\
\langle H_{I} \rangle&=&i g\,Z_{I}\,\int dq\,q\,
\left(A(q) -A^{\star}(q)\right),
\eq
and,
\be
\langle H \rangle=\langle H_{0} \rangle +\langle H_{I} \rangle+
\lmz\,\left(\int dq\,
|A(q)|^{2} -\rhz^{2}\right).
\ee
 $Z_{0}$ and $Z_{I}$ are functions of $\rhz$, to be determined in the
next section. Here, we have imposed a constraint on $A(q)$, which
we then integrate over. This introduces two new parameters 
$\lmz$ and $\rhz$, which prove very useful in simplifying the
calculations. 
Also, taking advantage of translation invariance in $\sigma$,
 all of the variational parameters are taken to
be independent of this variable.  To save writing, $\rho$ and
$\lambda$ are replaced by their ground state expectation values
$\rhz$ and $\lmz$, to be determined later.

The variational equation,
\be
\frac{\delta\langle H \rangle}{\delta A^{\star}(q)}
=Z_{0}\,q^{2}\,A(q)+\lmz\,A(q)-i g\,Z_{I}\,q=0,
\ee
has the solution
\be
A(q)=\eyz=i g\,Z_{I}\,\frac{q}{Z_{0}\,q^{2}+\lmz}.
\ee
 From
\be
\int dq\,|\eyz|^{2}=\rhz^{2}
\ee
it follows that
\be
\lmz=\frac{\pi^{2}}{4}\,g^{4}\,Z_{I}^{4}\,Z_{0}^{-3}\,\rhz^{-4}.
\ee
Both $\eyz$ and $\lmz$ being fixed, the only variables left are $a$ and
$\rhz$.
Later, we are going to expand $A$ around $\eyz$ (5.4). It is
important to notice that such an expansion always starts with
quadratic terms. All linear terms, in particular the interaction term,
are eliminated.
 
If we now try to compute $\langle H \rangle$ using (5.4),
 we find that the integral
over $q$ is linearly divergent. This is due to the translation invariance
in $q$: in a translationally invariant dynamical problem, energy is
proportional to the volume, so the finite quantity  is the energy
density. The solution for $\eyz$ may not look translationally 
invariant, but for large $|q|$, $q\,\eyz$ goes to a constant:
\be
q\,\eyz\rightarrow ig\,Z_{I}/Z_{0}.
\ee
As a result, if we put the system in a  one dimensional box of
size $L$, the leading term in
$\langle H \rangle$ is proportional to $L$ for large $L$,
 and the energy density $E$ is given by
\be
E=\frac{\langle H \rangle}{L}\rightarrow -g^{2}\,\frac{Z^{2}_{I}}{Z_{0}}.
\ee

We note that $E$ is negative. This is the sign of a phase transition:
The energy of this new state is lower than the energy of the former
ground state , the vacuum, which is zero. Of course only the energy
differences matter: We can add a constant and set the energy of the
new state equal to zero, and the energy of the vacuum would then be positive.

Ultimately, we are interested in  fluctuations local in $q$ around the 
translationally invariant ground state. These are generated
 by fluctuations in $\eyz$ that vanish rapidly for large $|q|$.
If we set
\be
A(q)=\eyz+\nabla A(q),
\ee
where the fluctuation term $\nabla A(q)$  vanishes for large $|q|$,
$H$ can be expanded to second order as
\be
H\rightarrow \langle H \rangle+ \int dq\,Z_{0}\,(q^{2}+\mzs)\,
(\nabla A(q))^{2}.
\ee
Here, we have defined a mass parameter by
\be
\mzs=\frac{\lmz}{Z_{0}}.
\ee

From the structure of the fluctuation term, it is clear that the
originally massless theory now has a mass. In fact, this could already
been seen by writing $\eyz$ as
\be
\eyz= ig\,\frac{Z_{I}}{Z_{0}}\,\frac{q}{q^{2}+\mzs}.
\ee
However, so far we have only solved the variational equation for $A(q)$.
It is possible that the solution of variational equation for 
$\rhz$ leads to a vanishing mass. To find the solution to this
equation, we first need to compute $N$, 
$Z_{0}$ and $Z_{I}$  as functions of $\rhz$.  We do this in the next
section by solving the recursion relation (4.2).

\vskip 9pt

\section{Solution Of The Recursion Relation For The Variational States}

\vskip 9pt

We start with the normalization constant for the variational state
$|s \rangle$ (4.6):
\be
N=\langle s|s \rangle=\sum_{n=2}^{\infty}
\langle n,\sigma=1|n,\sigma=1 \rangle=\sum_{n=2}^{\infty}
N(n,\sigma=1),
\ee
where,
$$
N(n,\sigma)=\langle n,\sigma|n,\sigma \rangle.
$$

The recursion relation (4.2) for the auxiliary states can be
rewritten as
\be
|n+1,\sigma \rangle =\int dq\,\int_{0}^{\sigma} d\sigma'
\,(\sigma -\sigma')^{a/2}\,\eyz\,\phi^{\dagger}(\sigma,q)
|n,\sigma' \rangle,
\ee
and the corresponding recursion relation for $N(n,\sigma)$ is
\be
N(n+1,\sigma)=\rhz^{2}\,\int_{0}^{\sigma} d\sigma'\,
(\sigma -\sigma')^{a}\,N(n,\sigma').
\ee
By either Fourier or Laplace transforming in the variable $\sigma$,
this is reduced to an algebraic equation, which is easily solved.
The result can be written as
\be
N(n,\sigma)=\rhz^{2 n}\,\sigma^{n(1+a)-1}\,
\frac{(\Gamma(1+a))^{n}}{\Gamma(n(a+1)},
\ee
and therefore,
\be
N=\sum_{n=2}^{\infty}\rhz^{2 n}\,\frac{(\Gamma(1+a))^{n}}{\Gamma(n(a+1)},
\ee
where eq.(6.1) has been used.

In deriving this result, we have imposed periodic boundary conditions
at the end points $\sigma=0$ and $\sigma=1$. This is a convenient
choice, although it will become clear later that the
precise form of the boundary
conditions are not important, so long as translation invariance in the
$\sigma$ variable is not violated. In the end, it is only the bulk
that matters.

Next, we will consider the matrix elements of $H_{0}$, which can
be written as
\be 
\langle s|H_{0}|s \rangle=M_{0}(\rhz)\,\int dq\,q^{2}\,|\eyz|^{2},
\ee
with,
\be
Z_{0}=\frac{M_{0}}{N}.
\ee

$M_{0}$ is then given by
\be
M_{0}=\rhz^{2}\,
\int_{0}^{1} d\sigma' \int_{0}^{\sigma'}
 d\sigma\,\langle s|\frac{
\mathcal{E}(\sigma,\sigma')}{\sigma' -\sigma} |s \rangle.
\ee
 Referring to eq.(6.3), the calculation of the right hand side
of this equation
 is similar to the calculation of $N$, except for the factor
$1/(\sigma' -\sigma)$ multiplying $\mathcal{E}(\sigma,\sigma')$. This
factor can be inserted in the state $|n,1 \rangle$ in 
$n+1$ different ways, and the resulting
$n+1$ terms are all identical because of the periodic boundary conditions.
We therefore compute the term say for $\sigma'=1$ and multiply the result
by $n+1$. Combining these observations, we have 
\be
M_{0}=\int_{0}^{1}
d\sigma\,(1-\sigma)^{a-1}\,\sum_{n=1}^{\infty}\,(n+1)\,N(n,\sigma),
\ee
and substituting for $N(n,\sigma)$ from (6.4)
 and doing the integral over $\sigma$
gives
\be
M_{0}=\sum_{n=1}^{\infty}\,(n+1)\,\rhz^{2 n}\,\frac{\Gamma(a)\,
(\Gamma(1+a))^{n}}{\Gamma(a+ n(1+a))}.
\ee

Next, we evaluate the matrix elements of the interaction Hamiltonian:
\be
\langle s|H_{I}|s \rangle=
\sum_{n=2}^{\infty}\,\langle n,1|H_{I}|n+1,1 \rangle+H.C.
\ee
From the structure of the vertex (2.2) (Fig.2), this calculation involves the
integral
\bq
Int.&=&\int_{\sigma_{1}}^{\sigma_{3}}
d\sigma_{2}\,\left((\sigma_{2}-\sigma_{1})\,(\sigma_{3} -\sigma_{2})\,
(\sigma_{3} -\sigma_{1})\right)^{(a-1)/2}\nonumber\\
&\times& \left(1+\frac{\sigma_{3}-\sigma_{2}}{\sigma_{2}-\sigma_{1}}
+\frac{\sigma_{2}-\sigma_{1}}{\sigma_{3}-\sigma_{2}}\right)\nonumber\\
&=&T\,(\sigma_{3} -\sigma_{1})^{a' -1} ,
\eq
where $a'=(3a+1)/2$ and,
\be
T=\frac{3a+1}{a-1}\,\frac{(\Gamma((a+1)/2))^{2}}{\Gamma(a+1)}.
\ee
 The matrix elemet of $H_{I}$ can now be written
as
\be
\langle s|H_{I}|s \rangle=ig\,T\,M_{I}\,\rhz^{2}\,\int dq\,q\,
\left(\eyz -\eyzs\right),
\ee
where,
\bq
M_{I}&=&\int_{0}^{1} d\sigma\,\sigma^{a'-1}\,\sum_{n=1}^{\infty}
(n+1)\,N(n,\sigma),\nonumber\\
&=&\sum_{n=1}^{\infty} (n+1)\,\rhz^{2 n}\,\frac{\Gamma(a')\,
(\Gamma(a+1))^{n}}{\Gamma(a'+n (a+1))}.
\eq

We note that eq.(6.16) is almost the same as eq.(6.9), except that the
exponent of $\sigma$ is now $a'-1$, instead of $a-1$. With this
replacement, the derivation of (6.16) goes through exactly as in
the case of (6.5). Putting together eqs.(5.1) and (6.15), we have,
\be
Z_{I}=N^{-1}\,\rhz^{2}\,T\,M_{I}.
\ee

We also note that $T$ and therefore $Z_{I}$ has a factor $1/(a-1)$,
so $a$ has to be greater
than one to have a finite result. This explains why the correlation
function (4.4) was needed. In the next section, we will use these
results to fix the values of $a$ and $\rhz$ for the ground state.

\vskip 9pt

\section{The Ground State}

\vskip 9pt

Let us first consider the dependence of $E$ on $a$ for fixed
$\rhz$. Because of the factor $1/(a-1)$ in the definition of
$T$ (6.14), as $a\rightarrow 1$,
\be
E\rightarrow -\frac{1}{(a-1)^{2}}\rightarrow -\infty.
\ee
Therefore, the ground state corresponds to the limit
$a\rightarrow 1$. We will set $a-1$, except, of course, in the
factor  $1/(a-1)$, and search for a minimum in the variable
$\rhz$. At $a=1$, 
the series for various quantities of interest  can be summed 
in closed form: 
\bq
N&=&\rhz\,\sinh(\rhz) -\rhz^{2},\nonumber\\
Z_{0}&=&\frac{\cosh(\rhz) -1+(\rhz/2)\,\sinh(\rhz)}
{\rhz\,\sinh(\rhz) -\rhz^{2}},\nonumber\\
Z_{I}&=&\frac{2}{a-1}\,\frac{\rhz\,\cosh(\rhz)- 2 \rhz+\sinh(\rhz)}
{\sinh(\rhz) -\rhz}.
\eq

Now consider the limit $\rhz\rightarrow \infty$,
\be
Z_{0}\rightarrow \frac{1}{2},\,\,Z_{I}\rightarrow \frac{2 \rhz}{a-1},
\,\, E\rightarrow - g^{2}\,\frac{\rhz^{2}}{(a-1)^{2}},
\ee 
 Clearly,  $E\rightarrow -\infty$  as $\rhz\rightarrow \infty$.
Therefore, the minimum of $E$, $E\rightarrow -\infty$, is
reached as $a\rightarrow 1$ and $\rhz \rightarrow \infty$, and
 the ground state values of  the two variational parameters
are fixed at $a=1$ and $\rhz\rightarrow \infty$.
However, this specification may appear somewhat problematic,
since it involves asymptotic limits tending to infinity. 
 We will now investigate the ground state in more detail, and
show that, with $m$ as the order parameter instead of $\rho$,
the ground state will be at some finite $m=m_{0}$.

We start by repeating an earlier observation: Only energy differences
are important; so we can adjust the ground state energy to be zero, 
then the energy of the vacuum and the excitations around it will go
to infinity. They will then effectively disappear from the spectrum.
It is now important to show that the energies of the 
local excitations around the ground state are finite. In particular,
 the generated mass (5.11),
\be
\mzs=\frac{\pi^{2}}{4}\,\left(\frac{g\,Z_{I}}{\rhz\,Z_{0}}\right)^{4}
\ee
is finite in the limit of infinite $\rhz$. From the asymptotic limits
(7.3) for $Z_{0}$ and $Z_{I}$,
\be
\mzs\rightarrow 64\,\pi^{2}\,\frac{g^{4}}{(a-1)^{4}},
\ee
so, in this limit, $\mzs$ is independent of $\rhz$. Therefore, instead
of $\rho$, if we take
$$
m^{2}=\frac{\lambda}{Z_{0}}
$$
as the order parameter, the limit $\rhz\rightarrow \infty$ can be
replaced by $m^{2}=\mzs$.

From eq.(7.5), it looks like $\mzs$ is singular at $a=1$. However, this
is if $g$ is held fixed in this limit. Instead, if we replace $g$
by $\mzs$ through
\be
g^{2}=\frac{(a-1)^{2}\,m_{0}}{8\,\pi},
\ee
 the singularity at $a=1$ 
disappears. Are we allowed to do this? After all, there is
no adjustable coupling constant in this problem; $g$ is there merely
to provide the correct dimension for the energy. If we let,
instead of $g$, $\mzs$ provide this dimension, the $a\rightarrow 1$
 limit can be taken without any blowup. 

What does the ground state look like? We will now show that the
world sheet is covered by a network of infinitely dense solid lines,
corresponding to an infinitely dense set of light cone graphs. We
start by noticing that in the series for $N$ (6.5), the summation
index $n$ corresponds to the number of solid lines. The n'th term in
this series is proportional to the relative probability for the
occurence of n lines. Denoting by $\langle n \rangle$ the average
value of $n$ weighed by these probabilities, we have,
\be
\langle n \rangle=\frac{1}{2 N}\,\rhz\,\frac{\partial N}
{\partial \rhz}.
\ee
Substituting  the large $\rhz$ limit (7.2) for $N$ gives
\be
\langle n \rangle\rightarrow  \frac{\rhz}{2}
\rightarrow \infty,
\ee
as $\rhz\rightarrow \infty$. This then shows that the ground state
 corresponds to an infinitely dense set of graphs on the
world sheet. It is then quite plausible that these graphs will
form a string of some kind [9,10]. In the next section, we will show that
this is indeed what happens.

What is the reason for the condensation of the world sheet graphs? It
is easy to see from the Hamiltonian that the lines of a graph actually
repel each other, so where does the attraction that causes
condensation come from? This attraction is provided by entroppy: The
entropy of a graph increases exponentially with the number of lines.
What we have here is the standard picture for phase transition;
entropy balances short range repulsion.

\vskip 9pt

\section{String Formation}

\vskip 9pt

To recapitulate, the ground state corresponds to the configuration
\be
i \phi_{i}=\eyz,\,\,\phi_{r}=0,\,\,a=1,\,\,m^{2} =\mzs,
\ee
where $\phi_{r,i}$ are the real and imaginay parts of $\phi$, 
and $\eyz$ is given by eq.(5.12). We will now consider 
time dependent fluctuations 
around this static configuration. The particular fluctuation that leads to
string formation corresponds to
shifting the momentum $q$ by the fluctuating field 
$\vts$. We therefore start by letting
\be
i \phi_{i}\rightarrow A_{0}(q+v(\tau,\sigma)),
\ee
 In addition, $\phi_{r}$ is taken to be non-zero, and with $q$ again
 shifted by $\vts$:
\be
\phi_{r}\rightarrow \phi_{r}(\tau,\sigma,q+\vts).
\ee
We will see later that a non-zero $\phi_{r}$ is needed to have
the correct canonical quantization of the fields.

$\eyz$ originally broke translation invariance in $q$, since it was
localized around $q=0$. The introduction of the
collective coordinate $v(\tau,\sigma)$ restores it, since
$$
q\rightarrow q+r
$$
will be accompanied by
$$
v\rightarrow v-r.
$$
$v$ is then the Goldstone mode of the  symmetry generated by
translations in $q$.
It will also turn out to be the string coordinate. In this
paper, we will only consider fluctuations generated by $v$,
with $a$ and $m^{2}$ fixed at their ground state values
given by (8.1).
Fluctuations in these parameters will hopefully be
studied in a future work.

If the ansatz given by (8.2) and (8.3) for $\phi_{i}$ and $\phi_{r}$ are
substituted in the kinetic energy term in the action (3.6), this term
 becomes,
\bq
K.E.&=& -2 \int d\tau \int d\sigma \int dq\,\phi_{r}(\tau,\sigma,
q+\vts)\,\partial_{\tau} \phi_{i}(\tau,\sigma,q+\vts)\nonumber\\
&\rightarrow& 2 i\, \int d\tau \int d\sigma \int
dq\,\phi_{r}(\tau,\sigma,q+\vts)\,\partial_{\tau}
A_{0}(q+\vts)\nonumber\\
&=& -2 g\,\frac{Z_{I}}{Z_{0}}\,
\int d\tau \int d\sigma \int dq\,\phi_{r}(\tau,\sigma,q)\,
\partial_{\tau}\vts\,
\partial_{q}\left(\frac{q}{q^{2}+\mzs}\right).\nonumber\\
&&
\eq
Here we have an action first order in the time $(\tau)$ variable,
with $\phi_{r}$ and $v$ as conjugate canonical variables.
This was the reason for keeping a non-zero $\phi_{r}$. Later,
$\phi_{r}$ will be eliminated using its equations of motion,
and resulting action will depend only on $v$.

Next we consider the fluctuations of $\langle H \rangle$,
which all come from $\langle H_{0} \rangle$. As explained earlier,
the expansion of $A(q)$ around $\eyz$ eliminates the interaction
term. We then make the replacement given by (8.2) and (8.3), and then
change the variable of integration from $q$ to $q- \vts$. The result is
\bq
\langle H \rangle &\rightarrow& \frac{1}{2} \int d\sigma \int d\sigma'
\int d q \int d q'\,\langle \frac{\mathcal{E}(\sigma,\sigma')}
{\sigma' -\sigma} \rangle\,\left(q -q'+v(\sigma')-
v(\sigma)\right)^{2}\nonumber\\
&\times& \phi^{\dagger} \phi(\sigma,q)\,\phi^{\dagger}
\phi(\sigma',q')
+ \int d\sigma\,\lambda(\sigma)\left(\int dq\,\phi^{\dagger}
\phi(\sigma,q) -\rho^{2}\right).
\eq

Expanding in powers of $q$ and $q'$, terms linear in $q$ and $q'$
involve the integral 
$$
\int dq\,q\,\phi^{\dagger} \phi(\sigma,q)=0,
$$
which vanishes
because of the symmetry (3.10). We can therefore set,
\be
\langle H \rangle= \langle \tilde{H} \rangle +\langle H_{v} \rangle,
\ee
where $\langle \tilde{H} \rangle$ is $v$ independent and 
$\langle  H_{v} \rangle$ is quadratic in $v$:
\bq
 \langle H_{v} \rangle &=&\frac{1}{2} \int d\sigma \int d\sigma'
\int d q \int d q'\,\langle \frac{\mathcal{E}(\sigma,\sigma')}
{\sigma' -\sigma} \rangle \,(v(\sigma) - v(\sigma'))^{2} \,
\phi^{\dagger} \phi(\sigma,q)\,\phi^{\dagger}
\phi(\sigma',q')\nonumber\\
&=& \frac{1}{2}\,\rhz^{4}\, \int d\sigma \int d\sigma'\,\langle 
\frac{\mathcal{E}(\sigma,\sigma')}{\sigma' -\sigma} \rangle \,
(v(\sigma) - v(\sigma'))^{2} .
\eq

We now evaluate the right hand side of this equation at
 $a=1$ and in the limit
$\rhz\rightarrow \infty$. First, setting $a=1$:
\bq
\langle s|\frac{\mathcal{E}(\sigma,\sigma')}{\sigma'-\sigma}|s \rangle
&=&\frac{1}{\rhz^{2}}\,\sum_{n=1}^{\infty} N(n,
1-\sigma'+\sigma)\nonumber\\
&=&\sum_{n=1}^{\infty} \rhz^{2 n-2}\,\frac{(1-\sigma' +\sigma)^{2 n -1}}
{\Gamma(2 n)}\nonumber\\
&=&\frac{1}{\rhz}\,\sinh(\rhz (1-\sigma'+\sigma)),
\eq
and then dividing by $N$ and taking the limit $\rhz\rightarrow
\infty$,
\bq
\langle \frac{\mathcal{E}(\sigma,\sigma')}{\sigma'-\sigma}\rangle
&=&\frac{\sinh(\rhz (1-\sigma'+\sigma))}{\rhz^{2}\,\sinh(\rhz)
- \rhz^{3}}\nonumber\\
&\rightarrow&\frac{1}{\rhz^{2}}\,\exp(- \rhz (\sigma' -\sigma)).
\eq

Notice that, in the limit $\rhz\rightarrow \infty$,
 the exponential function is sharply peaked at 
$\sigma' -\sigma=0$. To leading order, all the contribution to the
integral comes from the region $\sigma' -\sigma\approx 0$, and 
 we can therefore expand
$v(\sigma') -v(\sigma)$ in powers of $\sigma' -\sigma$ and keep only
 the first term:
$$
v(\sigma') -v(\sigma)\approx (\sigma' -\sigma)
 \partial_{\sigma} v(\sigma),
$$
with the result,
\bq
\langle H_{v} \rangle &\rightarrow &\frac{\rhz^{2}}{2}\,
\int_{0}^{1} d\sigma'\int_{0}^{\sigma'} d\sigma\,(\sigma'
-\sigma)^{2}\,\exp(\rhz (\sigma' -\sigma))\,
(\partial_{\sigma} v(\sigma))^{2}\nonumber\\
&\rightarrow& \frac{1}{\rhz}\,\int_{0}^{1} d\sigma\,
(\partial_{\sigma} v(\sigma))^{2}.
\eq

Notice that the dependence of $\langle H_{v} \rangle$ on $v$
is local in $\sigma$. This is in contrast to the original action
, which is non-local in $\sigma$. This localization is due
to the limit $\rhz\rightarrow \infty$. In this limit, the density
of the lines on the world sheet goes to $\infty$, and the distance
between two adjacent lines goes to zero (see end of section 7).
 Therefore, the interaction
between the world sheet lines becomes local.

Next, we have to compute $\langle \tilde{H} \rangle$ in the same
limit of the parameters. $\langle \tilde{H} \rangle$ is given by
(8.5), with $v=0$:
\bq
\langle \tilde{H} \rangle 
&=& \int d\sigma' \int d\sigma\,\langle \frac{\mathcal{E}(\sigma,
  \sigma')}{\sigma' -\sigma} \rangle\,\rhz^{2}\, \int dq\,
q^{2} \, \phi^{\dagger} \phi(\sigma,q)\nonumber\\
&+&
\int d\sigma \, \lmz \,\left(\int dq \,\phi^{\dagger}
\phi(\sigma,q) \,-\rhz^{2}(\sigma) \right).
\eq

 Now, in the integral
$$
\int dq\,q^{2}\, \phi^{\dagger} \phi(\sigma,q)= \int dq \,q^{2} \,
\left(|\eyz|^{2} + \phi_{r}^{2} (\sigma,q)\right),
$$
 the first term on the right was already included 
in the ground state energy, so it can be dropped, and so,
\bq
\langle \tilde{H} \rangle &\rightarrow&
 \int d\sigma' \int d\sigma\,\langle \frac{\mathcal{E}(\sigma,
  \sigma')}{\sigma' -\sigma} \rangle\,\rho^{2}(\sigma')\, \int dq\,
q^{2} \,\phi_{r}^{2} (\sigma,q)\nonumber\\
&+& \lmz \,\int d\sigma \int dq\,\phi_{r}^{2} (\sigma,q).
\eq
From eqs.(8.8) and (8.9),
\bq
&& \int d\sigma' \int d\sigma\,\langle \frac{\mathcal{E}(\sigma,
  \sigma')}{\sigma' -\sigma} \rangle\,\rho^{2}(\sigma')\, \int dq\,
q^{2} \,\phi_{r}^{2} (\sigma,q)\nonumber\\
&\rightarrow&\frac{1}{\rhz}\,\int d\sigma\,q^{2}\,\phi_{r}^{2}
(\sigma,q),
\eq
and therefore,
\be
\langle \tilde{H} \rangle \rightarrow \int d\sigma \int dq \,
(\frac{q^{2}}{\rhz} +\lmz)\,\phi_{r}^{2} (\sigma,q).
\ee

The total action is the sum of (8.4),(8.10) and (8.14). 
The dependence on $\phi_{r}$ in this action can  be eliminated using its
equations of motion:
$$
\phi_{r}= -g\,\frac{Z_{I}\,\rhz}{Z_{0}\,(q^{2} +\lmz\,\rhz)}
\,\partial_{\tau}\vts\,\partial_{q}\left(\frac{q}{q^{2}+\mzs}\right),
$$
and substituting in the action, and expressing $Z_{I}$, $Z_{0}$,
 $g$ and $\lmz$ in terms of $\rhz$ and $m_{0}$, 
\be
S = \int d\tau \int d\sigma\, \left(\frac{\rhz^{2} }
{m_{0}^{4} }\,(\partial_{\tau} v)^{2} -\frac{1}{\rhz}\,
(\partial_{\sigma} v)^{2} \right).
\ee

Apart from the normalization factors in front of the two terms,
 this is the standard string action in the light cone 
picture. However, these factors present a problem, since they are
singular in the limit  $\rhz\rightarrow \infty$.
This singularity can be eliminated by a rescaling of the sigma coordinate
by
\be
\sigma=(\rhz)^{- 3/2}\,\bar{\sigma},
\ee
and by defining
\be
\vts=\frac{m_{0}^{2}}{ \sqrt{2}\,\rhz}\,w(\tau, \bar{\sigma}).
\ee
In terms of these new variables, the action becomes,
\be
S= \int d\tau \int d\bar{\sigma}\,\left(\frac{1}{2}\,(\partial_{\tau} w)^{2}
- \frac{\mz^{4}}{2}\,(\partial_{\bar{\sigma}} w)^{2} \right).
\ee

Now consider the limits of integration in the new coordinate
$\bsi$. The limits in $\sigma$ were from 0 to 1, but for the present
purpose, it is better to start with
$$
-\frac{1}{2}\leq \sigma \leq \frac{1}{2}.
$$
Then the limits of $\bsi$ are 
\be
-\frac{1}{2}\,(\rhz)^{3/2}\leq \bsi \leq \frac{1}{2}\,(\rhz)^{3/2}
\rightarrow -\infty \leq \bsi \leq +\infty.
\ee

 What we have here is
an action that describes the bulk of an infinitely long
string. The boundary conditions are arbitrary, so long as
translation invariance in $\bsi$ is preserved.
 However, we know that the glueballs of $QCD$ are the
excitations of a closed string, and the action we have describes
the bulk of this string without specifying the boundary conditions.
We supply the boundary conditions as follows: We cut a piece of the
string of length $p^{+}$, where $p^{+}$ is the total light cone
momentum, end glue the ends together with periodic boundary
conditions. Because of translation invariance in $\bsi$, the location
of the cut does not matter. In the end, we have a string of length 
 $p^{+}$ with periodic boundary conditions.

From the foregoing, it may appear that $QCD\, 3$ is described by
 a standard string model. This, however, is an oversimplification:
Even in the context of a variational approximation scheme, we have
neclected the fluctuations of the variables $a$ and $\rho$. When these
are included say in a perturbative scheme, the result will be a
more complicated string model. We hope to investigate this model in a
future work.

\vskip 9pt

\section{Discussion}

\vskip 9pt

In this article, we have applied the world sheet approach to field 
theory developed in the earlier work to $QCD\,3$. The main tool is
the two dimensional field theory on the world sheet, which sums the
planar light cone graphs. We have investigated the ground state 
of this model using a variational trial state. Several intersting
results follow from this calculation. There is a phase transition
to a non-perturbative ground state of lower energy than the 
empty world sheet. This new ground state is generated by an
infinitely dense set of light cone graphs, suggesting an emerging
string picture. Also, in this new phase a finite mass is generated
in the originally massless model. Finally, we have studied a 
particular fluctuation in the background of the ground state
configuration, and showed a standard string in the light cone
framework emerges. However, this was done neglecting all other possible
fluctuations. A promising future direction of research is to study
other important fluctuations, and see how they modify the simple
string picture.

\vskip 9pt

\noindent{\bf Acknowledgement}

\vskip 9pt

This work was supported 
 by the Director, Office of Science,
 Office of High Energy  
 of the U.S. Department of Energy under Contract No.
DE-AC02-05CH11231.

\vskip 9pt

\noindent{\bf References}

\vskip 9pt

\begin{enumerate}

\item K.Bardakci, JHEP {\bf 1306} (2013) 066.
\item K.Bardakci, arXiv:1408.2556.
\item M.B.Halpern, Phys.Rev. {\bf D 16}, (1977) 1798; 
 I.Bars and F.Green, Nucl.Phys. {\bf B 148}, (1979) 445;
J.Greensite, Nucl.Phys. {\bf B 158}, (1979) 469;
M.Bauer and D.Z.Freedman, Nucl.Phys. {\bf B 450}, (1995) 209;
O.Ganor and J.Sonnenschein, Int. J. Mod.Phys. {\bf A 11}, (1996) 5701;
D.Karabali and V.P.Nair, Nucl.Phys. {\bf B 464}, (1996) 135;
Pys.Lett. {\bf B 379},(1996) 141;
D.Karabali, Chanju Kim and V.P.Nair, Nucl.Phys. {\bf B 524}, (1998) 661.
\item K.Bardakci and C.B.Thorn, Nucl.Phys. {\bf B 626}, (2002) 286.
\item C.B.Thorn, Nucl.Phys. {\bf B 637}, (2002) 272;
S.Gudmundsson, C.B.Thorn and T.A.Tran, Nucl.Phys. {\bf B 649},(2003)
3-38.
\item C.B.Thorn and T.A.Tran, Nucl.Phys. {\bf B 677},(2004) 289.
\item G.'t Hooft, Nucl.Phys. {\bf B 72},(1974) 461.
\item K.Bardakci, JHEP {\bf 0810}, (2008) 056.
\item H.P.Nielsen and P.Olesen, Phys.Lett. {\bf B 32}, (1970) 203.
\item B.Sakita and M.A.Virasoro, Phs.Rev.Lett. {\bf 24},(1970) 1146.

\end{enumerate}

\end{document}